\documentstyle[aps,prl]{revtex}
\input epsf
\begin{document}
\title{Spontaneous Magnetic Flux and Quantum Noise in a Doubly Connected
Mesoscopic  SND Junction}
\author{Alexandre M. Zagoskin and Masaki Oshikawa}
\address{Physics and Astronomy Department, The University of British Columbia,
6224 Agricultural Rd., Vancouver, B.C., V6T 1Z1, Canada}
\maketitle

\begin{abstract}
We consider spontaneous magnetic flux in a doubly connected  
s-wave/normal metal/d-wave superconductor (SND) Josephson junction.
The flux magnitude is not quantized and is determined by the
parameters of the system. In an isolated system, quantum fluctuations 
of the superconducting phase take place, leading to quantum flux noise.
 The possibilities of experimental
observations of the effect are discussed.
\end{abstract}

\vspace{0.1 in}

 Mesoscopic SNS systems containing superconductors with different pairing
 symmetry   promise a whole new range of phenomena, such as
half-periodic Josephson effect \cite{Zagoskin} and spontaneous currents
parallel to the interface \cite{avo}. In the latter paper, it was shown that 
due to local violation of $\cal T$-symmetry, spontaneous currents
should flow parallel to the SN interface in the normal part of the system;
the authors noted that such a current is difficult to observe directly,
since it will average to zero over short distances.

This difficulty does not appear in a doubly connected SND system. There
the spontaneous current would give rise to an equilibrium magnetic flux.
Its magnitude is  less than $\Phi_0$  
and is determined only by the parameters of the system.  It should not be mixed up with the fractional flux that is predicted in a contour with a $\cal T$-symmetry-violating junction \cite{avo,YipSunSouls,SigristBaileyLaughlin}. The latter is created by the current flow {\em across} the junction; the former is directly created by
 the spontaneous current  flowing {\em parallel} to the SN interface. 
It is therefore an analog of  persistent currents in normal mesoscopic rings
\cite{persistent}, when the $\cal T$-symmetry is   broken by an 
external magnetic flux.

\begin{figure}
\epsfysize=3 in
\epsffile{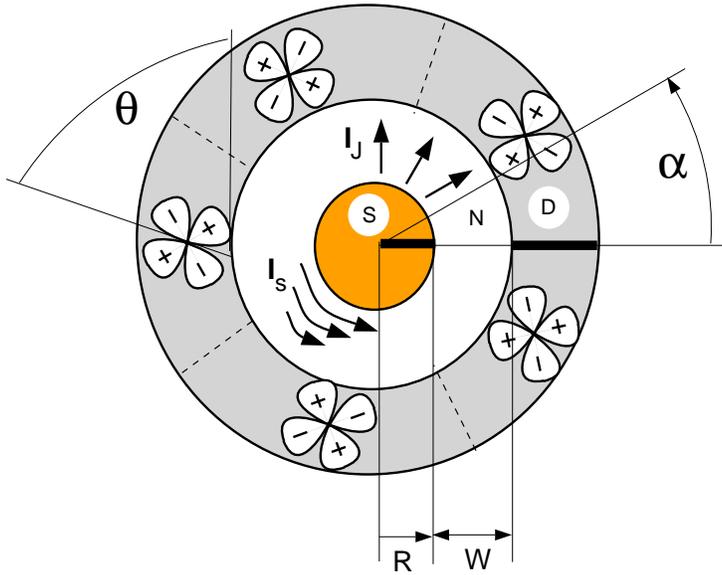}
\caption{Annular SND junction.  The c-axis of the d-wave superconductor is chosen to be
parallel to the SD boundary;  $\theta$ is the angle
between the SD boundary and the nodal plane of the d-wave order parameter
($0\leq\theta\leq\frac{\pi}{2}$).}\label{fig1}
 \end{figure}

 We consider here an annular SND junction (Fig.\ref{fig1}). 
   This idealised structure 
consists of the inner (s-wave) superconducting contact S, conducting ring N,
and the outer (d-wave) superconducting ring D. The splits in  S and D 
ensure that the screening supercurrents in these electrodes do not
mask the effects of spontaneous current in the normal part of the system, N.
 For the sake of simplicity,
we assume that
the orientation of the d-wave order parameter positive and 
negative lobes with respect to the 
radius vector is the same along  the d-wave ring. 

We  calculate  Josephson current in the system, following 
\cite{Zagoskin}. It is carried through the
normal region by two sets of Andreev levels, coupled to positive
and negative lobes of the d-wave order parameter (chosen to be real; 
the phase difference across the contact is thus $\phi$, where $\Delta_1e^{i\phi}$
is the order parameter of the s-wave superconductor). We are considering a
"small" junction limit, where the influence on the Josephson current of its own
magnetic field can be neglected. This imposes a condition (see Fig.1)
\begin{equation}
2\pi R < \lambda_J,
\end{equation}
where $\lambda_J = \sqrt{\hbar c^2/8\pi e j_c\Lambda}$ is the Josephson penetration length \cite{Barone}; $j_c$ is the critical  
Josephson current density,
and $\Lambda=W+\lambda_{L,1}+\lambda_{L,2}$ is the integral London penetration
length, $\lambda_{L,a}$ being the penetration length in the $a$th
superconductor.  In case of SNS junction 
\begin{equation}
\lambda_J \approx \lambda_F \sqrt{\frac{W}{\Lambda} 
\frac{c/v_F}{4\pi\alpha_{fs}}}
 <  \lambda_F\sqrt{ \frac{c/v_F}{4\pi\alpha_{fs}}} \equiv 
\lambda_F\zeta, \label{zeta}
\end{equation}
where $ \lambda_F$ is Fermi wavelength in the normal part of the system,
and  $\alpha_{fs} \equiv e^2/\hbar c \approx 1/137$.

Using the set of cylindric coordinates $(r,\alpha,z),$
we write the Bogoliubov-de Gennes equations in the normal region 
for a mode with vertical momentum $k_z$ as
\begin{eqnarray}
\left(\!\!\begin{array}{ll}
-\frac{1}{2m^*}\nabla^2-\mu_{\perp} & 0\\
0 & \frac{1}{2m^*}\nabla^2+\mu_{\perp}
\end{array}\!\!\right)
\left(\begin{array}{l}
u(r,\alpha;k_z)\\
v(r,\alpha;k_z)
\end{array}\right) = E\left(\!\!\begin{array}{l}
u(r,\alpha;k_z)\\
v(r,\alpha;k_z)
\end{array}\!\!\right), \end{eqnarray}
where $\mu_{\perp}=(k_F^2-k_z^2)/2m^* \equiv k_{\perp}^2/2m^*$, and 
$m^*$ is the effective mass of the electron. Expanding $u,v$ over asymuthal modes, $u(r,\alpha) = \sum_n U^{+}_n(r)e^{in\alpha}/\sqrt{r},$    $v(r,\alpha) = \sum_n U^{-}_n(r)e^{in\alpha}/\sqrt{r},$
we find:
\begin{equation}
-\frac{d^2}{dr^2}U^{\pm}_n(r)\!\!+\!\!\frac{n^2-1/4}{r^2}U^{\pm}_n(r) = 
(\pm 2m^*E\!+\!k_{\perp}^2)U^{\pm}_n(r).
\end{equation}

The Andreev levels are determined
from the quantization condition \cite{Kadigrobov}
\begin{eqnarray}
\oint p\:dr  = \int_{R}^{R+W}\!\!\!\!\!\!\!\!dr\:p(r,E) + \int_{R+W}^{R}\!\!\!\!\!\!\!\!dr\:p(r,-E) = 2\pi m + \pi \pm (\phi + \delta_{n,k_z}), \label{WKB}
\end{eqnarray}
where $m$ is integer,  and $p(r,E)=
(k_{\perp}^2
+
(n^2-1/4)/r^2+2m^*E)^{1/2}.$
The intrinsic phase shift 
$\delta_{n,k_z}$   in mode $(n,k_z)$ is zero if the corresponding lobe of d-wave order parameter carries positive sign, and $\pi$ otherwise \cite{Zagoskin}.
The radial  and asymuthal currents in a $n,k_z$-mode
are now easily found (cf.\cite{Zagoskin,avo}) as:
\begin{eqnarray}
J^{r(a)}_{n,k_z}(\phi+\delta_{n,k_z}) = j^{r(a)}_{n,k_z}  F(\frac{\pi}{\beta\epsilon(n,k_z)},\frac{1}{\tau\epsilon(n,k_z)};
\phi+\delta_{n,k_z}).
\label{JF}
  \end{eqnarray}
Here $j^{r,(a)}_{n,k_z}$ is the partial radial (asymuthal) current  
carried by Andreev levels with energies $E\approx 0$:
 \begin{eqnarray}
j^{r}_{n,k_z} = e\epsilon(n,k_z);\: j^a_{n,k_z} = \frac{e\epsilon(n,k_z) n}{2\pi(n^2-1/4){1/2}}
\left(\arctan\left[ \frac{(k_{\perp}(R+W))^2}{n^2 - 1/4} - 1\right] - 
\arctan\left[\frac{(k_{\perp}R)^2}{n^2 - 1/4} - 1 \right]\right); \label{ja}
\end{eqnarray}
and 
\begin{equation}\epsilon(n,k_z) =
\frac{k_{\perp}^2/2m^*}{[k_{\perp}^2(R+W)^2-(n^2-1/4)]^{1/2} -
[k_{\perp}^2R^2-(n^2-1/4)]^{1/2}}
\end{equation}
plays the role of the interlevel spacing $v_F^{\parallel}/2W$   of a planar  contact. The  currents  depend on the phase through
\begin{eqnarray}
  F(\frac{\pi}{\beta\epsilon(n,k_z)},\frac{1}{\tau\epsilon(n,k_z)};
\phi+\delta_{n,k_z}) 
=\frac{2}{\pi}
\sum_{m=1}^{\infty} \frac{(-1)^{m+1}\sin m(\phi+\delta_{n,k_z})}{\sinh \frac{\pi m}{\beta\epsilon(n,k_z)}}  \frac{\pi e^{-m/\tau\epsilon(n,k_z)} }{\beta\epsilon(n,k_z)},
\end{eqnarray}
where $\tau \gg \epsilon(n,k_z)^{-1}$ is the elastic scattering time. 
In the limit of zero temperature and no scattering 
$F(0,0;\phi)$ becomes the $2\pi$-periodic
sawtooth of unit amplitude.

First we find the Josephson (radial) current:
\begin{equation}
I_J(\phi) = \sum_{n,k_z}^{+} J_{r}(n,k_z;\phi)
+ \sum_{n,k_z}^{-} J_{r}(n,k_z;\phi+\pi).
\end{equation}
The "$\pm$"-sums are taken over zero- ($\delta_{n,k_z}=0$), and  $\pi$-levels ($\delta_{n,k_z}=\pi$), coupled to positive and negative lobes of the d-wave order parameter respectively \cite{Zagoskin}. At zero temperature and without
elastic scattering this  reduces to
\begin{equation}
I_J(\phi) = \frac{1+Z}{2}I_0 F(0,0;\phi) + \frac{1-Z}{2}I_0 F(0,0;\phi+\pi),
\end{equation}
where  $I_0 =  N_{\perp}e\bar{\epsilon}$ is the critical current in the system, 
$N_{\perp} \sim {\cal A}/\lambda_F^2$ is the number of quantum
conducting channels through the normal part of the system
(of crosssection area $\cal A$), and $\bar{\epsilon} \equiv \left<
\epsilon(n,k_z)\right> \sim v_F/2W$ is the average
interlevel spacing. The imbalance 
coefficient $Z$ ($|Z|\leq 1$) depends on the orientation of d-wave superconductor:
\begin{equation}
Z=\left\{\sum_{n,k_z}^{+}\epsilon(n,k_z)-\sum_{n,k_z}^{-}
\epsilon(n,k_z)\right\}\left/
\left\{\sum_{n,k_z}^{+}\epsilon(n,k_z)+\sum_{n,k_z}^{-}
\epsilon(n,k_z)\right\}\right.,
\end{equation}
and determines the equilibrium phase difference
across the junction, 
$\phi_0=\pm\frac{1-Z}{2}\pi.$
  The Josephson energy of the system is   thus
\begin{equation}
U(\phi) = \frac{1}{2e}\int d\phi I_J(\phi) = \frac{I_0}{2e}
\frac{(|\phi|-\phi_0)^2}{2\pi^2}\:\:\: (-\pi\leq\phi\leq\pi).
\end{equation}

In equilibrium, the total  Josephson current   is zero, because contributions from zero- and $\pi$-levels cancel. On the contrary, their contributions to the asymuthal current add up and lead to
finite spontaneous current. In the situation of Fig.\ref{fig1}, due to partial cancellation of terms with positive and negative $n$, 
it equals 
\begin{eqnarray}
I_s(\phi)=\left\{\begin{array}{ll}
\sum_{k_z}\sum_{n=n^-(\theta,k_z)+1}^{n^+(\theta,k_z)}
(J_a (|n|,k_z;\phi) - J_a (|n|,k_z;\phi+\pi))& {\rm if}\:\:
n^+(\theta,k_z)>n^-(\theta,k_z);\\
0 & {\rm if}\:\:
n^+(\theta,k_z)=n^-(\theta,k_z);\\
-\sum_{k_z}\sum_{n=n^+(\theta,k_z)+1}^{n^-(\theta,k_z)}
(J_a (|n|,k_z;\phi) - J_a (|n|,k_z;\phi+\pi))& {\rm if}\:\:
n^+(\theta,k_z)<n^-(\theta,k_z).
\end{array}
\right.
\end{eqnarray}
Here  ($[x]$ means   integral part of $x$)
\begin{eqnarray}
n^+(\theta,k_z)=\left[\frac{\tan\theta\sqrt{(k_{\perp}R_2)^2+1/4}}
{\sqrt{1+\tan^2\theta(k_{\perp}R_2)^2}} \right]; \:
n^-(\theta,k_z)=\left[\frac{\cot\theta\sqrt{(k_{\perp}R_2)^2+1/4}}
{\sqrt{1+\cot^2\theta(k_{\perp}R_2)^2}} \right].
\end{eqnarray}
In the limit $\beta,\tau,R,R+W\to\infty$ we return to the result \cite{avo}
for a planar junction \cite{note1},
\begin{equation}
I_s(\phi)=\frac{\pi e v_F}{\sqrt{2}W}\frac{{\cal A}}{\lambda_F^2}
\sin(\theta-\frac{\pi}{4}) (F(0,0;\phi)-F(0,0;\phi+\pi));\:\:\:0\leq\theta\leq \pi/2.
\label{25}
\end{equation}
 In the annular junction spontaneous current produces a spontaneous magnetic flux,
 $\Phi_s = \frac{{\cal L}}{c}I_s$,
 where ${\cal L} \sim \pi R$ is the self-inductance of the system. 
Flux magnitude can
be estimated as 
\begin{equation}
\Phi_s/\Phi_0 \approx \frac{1}{4\sqrt{2}} \frac{N_{\perp}}{\zeta^2}  
\frac{R}{W}.
 \end{equation}
  
\begin{figure}
\epsfysize=3 in
\epsfbox{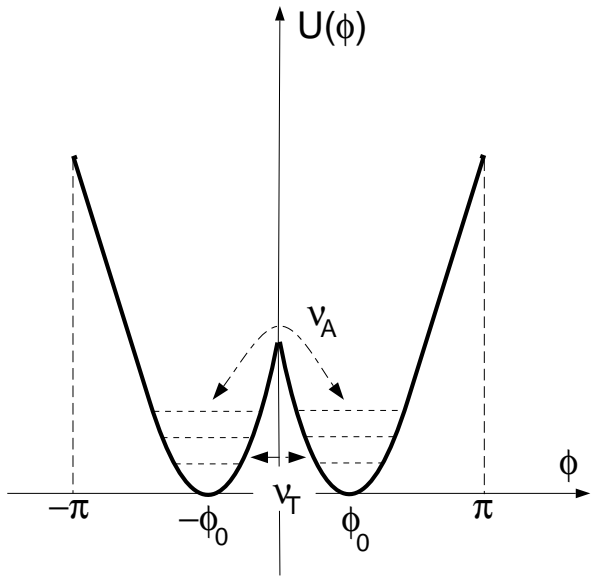}
\caption{Effective potential and transition rates for the phase variable
in an annular SND junction.}\label{fig4}
 \end{figure}

   The phase difference $\phi=\chi_1-\chi_2$ is  
canonically conjugate to 
 the operator of difference of Cooper pairs' numbers
 in the superconducting banks, $\Delta\hat{n} = 
\hat{n}_1-\hat{n}_2$\cite{Phi-N}. 
Therefore
in the presence of charging energy due to electron transfer
between the banks in an isolated junction (with classical capacitance $C$),
$\phi$ ceases to be a constant of motion.
 Its  evolution  can be mapped on the motion of a quantum 
particle in a one-dimensional ring
$-\pi\leq\phi<\pi$ (see e.g. \cite{Phi-N} and references therein), with
   the Hamiltonian 
\begin{eqnarray}
{\cal H}(\phi,\frac{\partial}{\partial\phi}) = 
\frac{\hbar^2}{2 M_Q}(\frac{1}{i}\frac{\partial}{\partial\phi})^2 + 
U(\phi);\:\:\:\:
 M_Q = \frac{C\hbar^2}{16 e^2} \equiv \frac{\hbar^2}
{8 \varepsilon_Q}.
\end{eqnarray}
The first term describes the charging energy.
   The effective potential $U(\phi)$  (Fig.\ref{fig4})
  corresponds
to the Josephson energy:
\begin{equation}
U(\phi) = \frac{\hbar I_0}{4\pi^2 e}(|\phi|-\phi_0)^2 =
\frac{N_{\perp} \bar{\epsilon}}{4\pi^2}(|\phi|-\phi_0)^2 = 
\frac{M_Q\omega_0^2}{2}(|\phi|-\phi_0)^2,
\end{equation}
where $\bar{\epsilon}$ was introduced earlier, 
 and
$\omega_0 = \sqrt{\left(\frac{C\hbar^2}{16 e^2}\right)^{-1}\cdot\frac{\hbar I_0}
{2\pi^2 e}} = \sqrt{\frac{8}{\pi^2} N_{\perp} \cdot \frac{\varepsilon_Q
\bar{\epsilon}}{\hbar^2}}
$
is the  frequency of small oscillations in a separate well. 
  The dynamics of the system is thus determined by four parameters:
 $\varepsilon_Q,$ $\bar{\epsilon},$ 
 $N_{\perp}$,
and   $\phi_0$. We will consider the case $|\phi_0|\ll\pi$
(or  $|\pi-\phi_0|\ll\pi$, in which case we should take $0\leq\phi<2\pi$). Then in the  limit  
$\hbar\omega_0, k_BT \ll U(0),$ we can use the two-well approximation
\cite{LLIII},
and  the transition rate between the wells will be given by
$\Gamma = \nu_A+\nu_T$, where the rate for thermally activated transitions
$\nu_A
\sim \omega_0 e^{-\beta U(0)}$, and for
quantum tunneling $\nu_T \sim \omega_0 e^{-\frac{U(0)}{\hbar\omega_0}}$.

The phase fluctuations  create finite voltage,
$V = \frac{\hbar}{2e}\dot{\phi},$, and therefore normal current
$GV$ flows in the normal region (its conductance    $G \sim N_{\perp}\frac{e^2}{\pi\hbar}$). The corresponding dissipative function 
 and the decay decrement are 
\begin{equation} 
{\cal F} = \frac{1}{2}\frac{d}{dt}E = \frac{1}{2}GV^2 = \frac{1}{2}G\left(\frac{\hbar}{2e}\right)^2\dot{\phi}^2;\:\:\: 
\gamma = \frac{1}{M_Q\dot{\phi}}\frac{\partial{\cal F}}{\partial\dot{\phi}} = 
\frac{G\hbar^2}{4 e^2 M_Q} = \frac{4}{\pi} N_{\perp} 
\frac{\varepsilon_Q}{\hbar}.
\end{equation}
 The character of quantum fluctuations of the magnetic flux  
drastically depends on dissipation.
In the limit $\hbar\omega_0 \ll U(0)$ 
the conclusions basically reduce to the following\cite{MQT}.
The tunneling rate exponent multiplies by $\sim (1+
\gamma/\omega_0)$. Therefore if $\gamma > \omega_0$,
tunneling is suppressed. (Strictly speaking, in this case the description in terms of initial two-well
potential becomes meaningless.) 
 If, on the other hand, 
\begin{equation}
\frac{\gamma}{\omega_0} = \sqrt{N_{\perp}\frac{\varepsilon_Q}{\bar{\epsilon}}}
\ll 1, \label{gamma<omega}\end{equation}
 quantum tunneling is possible,
but quantum coherence between the states in left and right wells is
 destroyed\cite{note2}.

The problem is simplified due to phase dependence of the spontaneous 
current (\ref{25}), which can be approximated by
\begin{equation}
\Phi(\phi) =  \Phi_s\: {\rm sgn}\: \phi ({\rm mod}\: 2\pi).
\end{equation}
 This
 allows to describe the dynamics of $\Phi$
in terms of a two-level system\cite{TLS}: state "$\downarrow$" is
when  $\phi$ is in the left well ($\Phi = -\Phi_s$), 
 state "$\uparrow$" when $\phi$ is in the right well ($\Phi = \Phi_s$).
Therefore we can write a set of coupled master equations  
for diagonal terms of density matrix (\cite{Blum}, Ch.7):
\begin{eqnarray}
\left\{
\begin{array}{l}
\frac{\partial}{\partial t}\rho_{\uparrow} = \Gamma (\rho_{\downarrow} - 
\rho_{\uparrow}) \\
\frac{\partial}{\partial t}\rho_{\downarrow} = -\Gamma (\rho_{\downarrow} - 
\rho_{\uparrow})
\end{array}\right. \label{telegraph}
\end{eqnarray}
Here $\rho_{\uparrow,\downarrow}$ gives the probability of occupation of right
(left) well (spontaneous flux $\pm\Phi_s$).
Eq.(\ref{telegraph}) describes random telegraph noise of spontaneous
magnetic flux, with autocorrelation
function and spectral density\cite{Gardiner}
\begin{eqnarray}
\left<\Phi(t)\Phi(0)\right> = \Phi_s^2 e^{-2\Gamma|t|};
\:\:
\left(\Phi^2\right)_{\omega} = \frac{4\Phi_s^2  \Gamma}{\omega^2 +
4 \Gamma^2}.
\end{eqnarray}
The noise is   temperature dependent  at $T > T^* = \hbar\omega_0/k_B$, where 
  thermally activated transitions are more prominent  than tunneling. 
Below $T^*$, the activated processe freeze out, and quantum tunneling remains
the sole source of flux noise.
Magnetic noise intensity detected at a certain frequency $\Omega$, as $\Gamma$
is changed with temperature, will peak at $\Gamma(T) = \Omega/2$.

Let us make some numerical estimates. Consider the case when the
role of the normal part of the system is played by 2D electron gas of a
GaAs-AlGaAs heterostructure, with $\lambda_F = 450$\AA ~and
$v_F = 2\times 10^7$ cm/sec. Then the number of transverse modes
is $N_{\perp} = 4\pi R/\lambda_F$, and the condition of "small" contact
(\ref{zeta}) becomes 
$N_{\perp} < 2\zeta \sqrt{W/\Lambda} \approx 250 \sqrt{W/\Lambda}.$
It will be satisfied if e.g. $N_{\perp} = 200,\:\:(R\approx
7000\:$\AA$\:),\:W = 1000\:$\AA.~The interlevel spacing is then 
$\bar{\epsilon} \approx 10^{-15}$erg. 
The corresponding 
spontaneous flux amplitude $\Phi_s \approx 0.02\Phi_0$ is within the
experimental capabilities.

The    applicability condition of the  two-well model   imposes the condition $\varepsilon_Q/\bar{\epsilon} \ll (N_{\perp}\phi_0^4)/128\pi^2$,
which will be satisfied if $
\phi_0 >   (128\pi^2/N_{\perp})^{1/4} N_{\perp}^{-1/4} \approx 0.42.
$

The weak dissipation condition
(\ref{gamma<omega}) translates into $\epsilon_Q \gg 5\times 10^{-18}$erg (that is, total capacitance of the system should exceed   $\sim 3\times 10^{-14}$F).
 This yields for $\hbar\omega_0 \ll \sqrt{\frac{8}{\pi^2}\bar{\epsilon}^2} \approx \bar{\epsilon},$
that is, $\omega_0 \ll 10^{12}$ sec$^{-1}$. The flux noise characteristic
frequency $\Gamma$ is 
exponentially  dependent on $\phi_0^2$ (through $U(0)$), and scales with 
temperature from $\nu_T \sim \omega_0 \exp[-U(0)/\hbar\omega_0]$ at $T\ll T^*$
to $\nu_T \sim \omega_0$ at $T\gg T^*$, and will probably reach a fraction
of GHz. 
 
In conclusion, we have found the Josephson current
and spontaneous magnetic flux 
in an  annular SND junction.   
The magnitude of the flux is not quantized; it depends on  
the parameters of the system and is of the order of   few percent
of $\Phi_0$ for a typical configuration. 

We demonstrated that
in an isolated system, due to charging effects, there will be quantum
noise of the spontaneous magnetic flux, with characteristic frequency
 strongly dependent on the configuration of the system and its surroundings.
It can reach a fraction of GHz. The noise intensity at a fixed frequency
will be a non-monotone function of temperature, allowing to measure
the transition rate
  $\Gamma(T)$ directly and thus separate the contributions from thermally activated processes and quantum tunneling.

We hope that such systems provide experimental observation of a novel dynamical effect in $d$-wave superconductivity.

We are thankful to I. Affleck, I. Herbut, and P. Stamp  for helpful discussions.
M.O. was supported by Killam Postdoctoral Fellowship.

\end{document}